\documentclass[12pt]{article}
\usepackage{graphics, color}
\usepackage{graphicx}
\usepackage{amssymb, amsmath, tensor}
\pdfoutput=1

\def\gsim{\, \rlap{$>$}{\lower 1.1ex\hbox{$\sim$}}\,}
\def\lsim{\, \rlap{$<$}{\lower 1.1ex\hbox{$\sim$}}\,}
\newcommand{\Op}{\mathcal{O}}

\renewcommand{\vec}[1]{\boldsymbol{#1}}

\begin{document}

%Title page

\begin{titlepage}
\bigskip
\bigskip\bigskip\bigskip
%\centerline{\Large Boundary Construction of the Maxwell- and Gravitational Fields}
\centerline{\Large Construction of Bulk Fields with Gauge Redundancy}
\bigskip\bigskip\bigskip
\bigskip\bigskip\bigskip

 \centerline{{\bf Idse Heemskerk}\footnote{\tt idse@physics.ucsb.edu}}
\medskip
\centerline{\em Department of Physics}
\centerline{\em University of California}
\centerline{\em Santa Barbara, CA 93106}
\bigskip
%  \centerline{{\bf Joseph Polchinski}\footnote{\tt joep@kitp.ucsb.edu}}
% \medskip
% \centerline{\em Kavli Institute for Theoretical Physics}
% \centerline{\em University of California}
% \centerline{\em Santa Barbara, CA 93106-4030}\bigskip
% \bigskip
%\centerline{\bf ...}
\bigskip
\bigskip\bigskip

%ABSTRACT

\begin{abstract}
We extend the construction of field operators in AdS as smeared single trace operators in the boundary CFT to gauge fields and gravity. Bulk field operators in a fixed gauge can be thought of as non-local gauge invariant observables. Non-local commutators result from the Gauss' law constraint, which for gravity implies a perturbative notion of holography. We work out these commutators in a generalized Coulomb gauge and obtain leading order smearing functions in radial gauge.
\end{abstract}
\end{titlepage}
\baselineskip = 16pt

%\tableofcontents

\section{Introduction}

In the supergravity limit of AdS/CFT, a large $N$, strong coupling gauge theory is dual to a theory of gravity which is local in the sense that it is described by a local Lagrange density.
The latter implies the canonical notion of locality that observables with spacelike separated support commute, so that no measurement can influence another measurement outside its future lightcone. 
Understanding bulk locality from the perspective of the lower dimensional boundary gauge theory is a challenge \cite{Gary:2009ae,HoloFromCFT}. A canonical approach is the $1/N$ perturbative construction of bulk field operators from smeared single trace boundary operators \cite{BDHM,Bena99,KLL2011,ourotherpaper}.  The only case that has been worked out so far is the scalar field, for which one can find smearing functions $K^{(n)}$ such that the following operator identity holds
\begin{align} 
\label{smearingscalar}
\phi(y) 
=\int d^dx_1\,& K^{(0)}(y|x_1){\cal O}(x_1)
\\\nonumber
& + \frac{g}{N} \!\int\! d^dx_1\,d^d x_2\, \sqrt{-g'} K^{(1)}(y|x_1, x_2)  {\cal O}(x_1) {\cal O}(x_2)  + O(1/N^2)\,.
\end{align}
In this paper we work out the cases of the gauge field and gravitation field. 
Technically, the construction is now complicated by the need to fix a gauge and solve the constraints before the field operators can be expressed in terms of gauge invariant boundary currents. Conceptually, one may wonder if the constructed field operators have physical meaning and if in the gravitational theory one can talk about locality in the sense of local commuting observables, given the mantra that there are no local diffeomorphism invariant observables.

We begin by reviewing that non-local gauge invariant operators can be traded for local operators in a fixed gauge, but that these operators can have local commutators only up to the non-locality required by the constraint. We then explicitly carry out the canonical quantization of the Maxwell field in AdS, in an appropriate generalization of Coulomb gauge. We work out the non-local commutators due to Gauss' law in detail and from mode expansions we obtain leading order smearing functions in Coulomb and radial gauge.
For the gravitational field we work out the Gauss' law commutators in transverse gauge, which in this case implies a perturbative notion of holography. Finally we obtain the leading order smearing function in Pfefferman-Graham gauge.
In the discussion we comment on future directions.

\section{Local observables and gauge invariance}

\begin{figure}
\begin{center}
\includegraphics[scale=1]{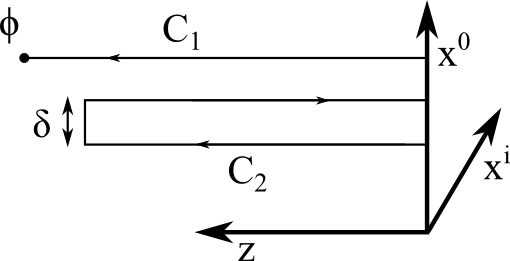}
\end{center}
\caption{A Wilson line along $C_1$ connects a charged scalar to the boundary. The gauge invariant integral of $A$ along $C_2$ reduces to the scalar potential $A_0$ in radial gauge.}
\label{wilsonlines}
\end{figure}

A charged scalar field field $\phi$ is not gauge invariant. However, if there is a place where the potential is fixed we can contruct a gauge invariant operator by connecting the field to this place with a Wilson line. For example, consider a half space that has a flat boundary on which the gauge field $A_\mu$ vanishes. Let $x^\mu$ be the coordinates on the boundary and $z$ a ``radial coordinate'' away from the boundary (see fig. \ref{wilsonlines}). Then one gauge invariant non-local operator is
\begin{equation}
e(z,x) \equiv \phi(z,x)\exp\left(i\int_{C_1} A\right) = \phi(z,x) \exp\left(i\int_{0}^z A_z(z',x) dz' \right).
\end{equation}
Another gauge invariant operator that we could construct is the Wilson
loop in figure \ref{wilsonlines}
\begin{equation}
V(z,x) \equiv \lim_{\delta\to 0} \frac{1}{\delta} \oint_{C_2}A = \int_0^z E_z(z,x) dz,
\end{equation}
which measures the potential difference between us and the boundary.
It is clear that 
$$[e(z,t,x^i), V(z',t,x^i)]\neq 0,$$ 
even though $(z,t,x^i)$ and $(z',t,x^i)$ are spacelike to each other because these non-local operators will overlap.
However, these operators are precisely $\phi(z,x)$ and $A_0(z,x)$ if we choose radial gauge $A_z=0$. From the gauge fixed perspective the non-locality in the commutator comes from having to impose the Gauss' law constraint. 
%This is most easily worked out in Coulomb gauge where the constraint fixes $A_0$ in terms of the charge density at equal time. 
As an illustration, consider flat space quantum electrodynamics in Coulomb gauge. The gauge field is fixed to go to zero at spacelike infinity and we can then define non-local gauge invariant operators \cite{Nguyen:1989iz} that have ``Wilson lines'' connecting the charged field to infinity or equivalently work in the gauge fixed theory \cite{BjorkenDrell}. Gauge fixing $\nabla \cdot \vec{A} = 0$ yields the constraint
$
-\nabla^2 A_0 = e\bar\psi \gamma^\mu \psi,
$
which after canonical quantization of $\psi$ leads to the non-local equal time commutator
\begin{equation} \label{QEDGauss}
[A_0(\vec{x},t), \psi(\vec{x}',t)] = -\frac{e}{4\pi |\vec{x}-\vec{x}'|} \psi(x',t),
\end{equation}
and consequently a non-local commutator between the electric field and the electron. 

For gravity, the situation is similar. It is clear that it is useful to talk about localized events even in gravitational theories, but that position should be defined in a coordinate independent, relative way. If space had a boundary, that boundary would be a convenient reference point. So we consider again half space with a flat boundary, where now the metric is held fixed (we have in mind asymptotically Poincare AdS with a cut-off\footnote{There is no essential difference if we take the cut-off to zero so that boundary becomes asymptotic boundary if we make the appropriate subtractions.}).
Then we can define a scalar field $\phi(z,x)$ in that space with 
\begin{equation} \label{coordinateconvention}
\ln z = \int \sqrt{-g_{MN}(y^L(s))\frac{dy^M}{ds}\frac{dy^N}{ds}} ds, \qquad y^M = (z, x^\mu) = (z,t,x^i)= (t, u^a),
\end{equation}
where $z$ measures the log of the geodesic distance along a geodesic that starts at $x$ normal to the boundary. This diffeomorphism invariant field operator is a non-local functional of $g_{MN}$. However, this is exactly the same thing as imposing the gauge fixing condition that the metric take the form 
\begin{equation} \label{PGmetric}
ds^2 = g_{MN}dy^Mdy^N = \frac{dz^2}{z^2} + \gamma_{\mu\nu}(z,x) dx^\mu dx^\nu, \qquad \lim_{z\to 0} z^2 \gamma_{\mu\nu}(z,x) = \eta_{\mu\nu},
\end{equation}
analogous to Gaussian normal coordinates \cite{WHRG} and then considering the diffeomorphism variant field $\phi(z,x)$ in these coordinates. 
One can also define gauge invariant operators that reduce to particular metric components in some gauge, namely the length of curves with two ends at the boundary.
Just like before those will have non-local commutators due to the constraint, which we will work out in section \ref{gravitysection}. 
An important difference between gravity and gauge theory is that Wilson loops in the bulk, which in the gauge theory are observables of compact support that commute with the electric field at infinity, have no gravitional analogue.
%, because to specify the set of points that constitute a loop will always require the boundary as a reference point (analogous to the way we previously defined a curve as the geodesic normal to the boundary of a certain proper length).

To summarize: for all intents and purposes we can talk about local fields after gauge fixing as physical observables, but these will have commutators that are local only to the extent allowed by the constraint. We will see this explicitly for the $U(1)$ gauge field and gravity in AdS.

\section{Maxwell field in AdS}

The Maxwell field coupled to a charged scalar in AdS$_{d+1}$ has action
\begin{equation}
S= -\int d^{d+1} y \sqrt{-g} \left[\frac{1}{4}  F_{MN}F^{MN}  + (D_M\phi)^* D^M\phi + m^2 \phi^*\phi \right], 
\end{equation} 
and the equations of motion for the gauge field are
\begin{align} \label{EOM}
\frac{1}{\sqrt{-g}}\partial_M\left(g^{MM'}g^{NN'}\sqrt{-g}(\partial_{M'} A_{N'}-\partial_{N'} A_{M'})\right) = &J^N, \\& J^N \equiv i q (\phi^* D^N \phi-D^N \phi^* \phi),
\end{align}
where $D_M$ denotes the AdS gauge covariant derivative $D_M \phi =
(\partial_M - iqA_M)\phi$ and $\nabla_M$ the covariant derivative with respect to the metric on the Poincar\'e patch, given by \eqref{PGmetric} with $h_{\mu\nu}=z^{-2}\eta_{\mu\nu}$. For later use we define the electric field, whose flux is conserved by Gauss' law (see \eqref{coordinateconvention} for index conventions)
\begin{equation} \label{electricfield}
E^a \equiv g^{ab}g^{00} F_{0b}, \qquad\qquad \frac{1}{\sqrt{-g}} \partial_a (\sqrt{-g} E^a) = J^0.
\end{equation}

The goal is to fix a gauge, solve the constraint, and canonically quantize the fields. We can then calculate the commutator analogous to \eqref{QEDGauss} and obtain the leading order smearing function. 
Radial gauge $A_z=0$ and Lorentz gauge $\nabla_M A^M=0$ are usually convenient gauges because they respectively preserve the full AdS and Poincar\'e symmetry of the metric, and we will obtain the smearing function in radial gauge at the end of this section. However, these gauge conditions make it harder to see that Gauss' law leads to non-local commutators. 
%solving the constraint because they leave $A_0$ and $A_a$ mixed\footnote{While we can solve the equations in both cases by splitting the field into appropriately defined transverse and longitudinal components, the scalar potential $A_0$ will not be related to the current at fixed time and therefore have a convoluted commutator with a charged field.}. 
We therefore generalize the Coulomb gauge condition, which in flat space is $\partial^i A_i = 0, $ and eliminates the term involving the spatial components from the constraint.
%, leaving us with an equation involving only $A_0$.
The generalization of this condition to AdS is not $\nabla^a A_a=0$. Instead we see from \eqref{EOM} that we can eliminate the spatial components from the constraint by imposing 
\begin{equation} \label{gaugefixingcondition}
-\partial_a\left(g^{ab}g^{00}\sqrt{-g} A_{b})\right) = \partial_a(z^{3-d}A_a)= 0.
\end{equation}
This reduces the constraint to
\begin{equation} \label{constraint}
z^{d+1} \partial_a (z^{-d-1} E^a) = -z^{d+1}\partial_z (z^{3-d} \partial_z A_0) - z^4 \partial_i^2 A_0 = J^0,
\end{equation}
and the remaining equations to
\begin{align} \label{spatialMaxwell}
\partial_a\left(z^{3-d}(\partial_{a} A_{d}-\partial_{d} A_{a})\right) - z^{3-d}\partial_0^2 A_{d}= &P_{dc} (z^{-d-1}J^c),
\end{align}
where $\partial_i^2 = \delta^{ij}\partial_i\partial_j$ and $P$ is a generalized transverse projector
\begin{align}
P_{ad} \equiv \delta_{ad} - z^{3-d}\partial_a (\partial_c  z^{3-d}\partial_c)^{-1} \partial_d.
\end{align}
The spatial equations can be decoupled using the gauge condition \eqref{gaugefixingcondition}, yielding 
\begin{align}\label{spatialMaxwelldecoupled}
\partial^2 A_z + \partial_z (z^{d-3}\partial_z (z^{3-d}A_z))  &=z^{d-3} P_{zc}( z^{-d-1} J^c ), 
\\
\partial^2 A_j +z^{d-3} \partial_z (z^{3-d}\partial_z A_i) &= z^{d-3} P_{ic} (z^{-d-1} J^c ),
\end{align}
where $\partial^2 = \eta^{\mu\nu}\partial_\mu \partial_\nu$. The gauge condition and resulting equations of motion are not $AdS_{d+1}$ covariant, nor are they covariant under the Euclidean $AdS_{d}$ symmetries of a constant timeslice. Green's functions for these equations will therefore not be functions of a single invariant distance.
With boundary conditions that we will discuss in the shortly, the constraint determines $A_0$ uniquely in terms of the charge density. 
We can now canonically quantize the remaining physical degrees of freedom subject to the gauge fixing condition,
\begin{align}
[ \hat A_a(u), \hat \Pi_b(u')] &= i\delta^{tr}_{ab}(u-u'), & \Pi_a =& z^{3-d}\partial_0 A_a,
\\
[\hat\phi(u',t),\hat\pi(u,t)] &= i\delta^d(u-u'), &\pi =& -\sqrt{-g}(D^0\phi)^*,
\end{align}
where the transverse delta generalizes the one in flatspace \cite{BjorkenDrell},
\begin{equation}
\delta_{ab}^{tr}(u-u') = z^{d-3} P_{ab} z^{3-d} \delta^{d}(u-u').
\end{equation}
The commutator between the current operator and the scalar field is then
\begin{equation} \label{currentcommutator}
[\hat J^0(u,t),\hat\phi(u',t)] = \frac{q}{\sqrt{-g}} \hat\phi(u) \delta^d(u-u').
\end{equation}
Gauge/gravity duality maps these operators in the bulk of AdS to operators in a boundary theory through the extrapolate dictionary \cite{Polchinski:2010hw}
\begin{equation} \label{dictionary}
\lim_{z\to 0}\sqrt{-g} \hat F^{z\mu}(z,x) = \hat j^\mu(x), \qquad  \lim_{z\to 0} z^{\Delta_+}\hat\phi(z,x) = \Op(x),
\end{equation}
The explicit solution for $A_0$ in terms of $J_0$ is slightly
different for the cases $d=1$ and $d>1$. In $d=1$ the Maxwell
equations are pure constraint and the solutions are simple. We will therefore solve it first.

\subsection{AdS$_2$}
In AdS$_2$, the Coulomb gauge condition is $\partial_z (z^2 A_z)=0$. We can consistently impose the slightly stronger condition $A_z=0$ (radial gauge), after which the equations take the form
\begin{align} \label{MaxwellAdS2}
z^2 \partial_z (z^2 \partial_z A_0) = -J^0, \qquad z^4 \partial_0 \partial_z A_0 = J^z,
\end{align}
with the homogeneous solution
\begin{equation}
A_0 = \frac{a}{z}+ b(t).
\end{equation}
The boundary conditions in $d=1$ are less intuitive than in higher dimensions. Spatial infinity consists of two disconnected points so flux can come in from $z=\infty$. We demand that there is not so $z^2\partial_z A_0|_{z=\infty}= a= 0$. On the other side, at $z=0$, we fix $b=0$ so that the unique homogeneous solution $A_0=0$. Solving the Gauss' law constraint \eqref{MaxwellAdS2} for a point charge a point charge $$J^0(z') = \delta(z''-z') z'^2$$ 
with these boundary conditions then yields the Coulomb potential
\begin{equation}
\Phi(z,z') = \Theta(z-z')\left( \frac{1}{z} - \frac{1}{z'} \right).
\end{equation}
The solution for an arbitrary localized charge distribution is
\begin{equation}
A_0(z) = \int \frac{dz'}{z'^2} \Phi(z,z') J^0(z',t).
%= -\int_z^\infty \frac{dz'}{z'^2} \left( \frac{1}{z'} - \frac{1}{z} \right) J^0(z',t)
\end{equation}
The second equation \eqref{MaxwellAdS2} is implied by current conservation. From \eqref{electricfield}, \eqref{currentcommutator} we then find the non-local commutator
\begin{equation} \label{AdS2commutator}
[\hat E(z), \hat\phi(z')] = q z^2\Theta(z-z')\hat\phi(z'),
\end{equation}
between the electric field and the charged scalar.
As explained in the previous section one way to understand this is that $\phi(z')$ is a secretly non-local gauge invariant operator that creates both the charged scalar and its (longitudinal) electric field. Alternatively we simply notice that \eqref{AdS2commutator} is required by charge conservation. Gauss' law relates the integral of electric flux through the boundary to the total charge in the bulk so that in $d=1$ the electric field is proportional to the total charge operator
\begin{equation}
\hat Q_{bulk} = \lim_{z\to 0} z^{-2} E^z.
\end{equation}
The extrapolate dictionary \eqref{dictionary} now just says $\hat Q_{bulk} = \hat Q_{boundary}$, as we might have expected.

\subsection{AdS$_{d+1>2}$}
The general case of AdS$_{d+1}$, while conceptually the same, is technically useful as a warmup for gravity. 
Fourier transforming along the spatial boundary directions, the homogeneous solution to the constraint \eqref{constraint} is
\begin{align}
%A_L^\mu &= a_L^\mu(k) z^{d-2} + b_L^\mu(k), & P^\mu_\nu a^\nu &= P^\mu_\nu b^\nu = 0,
%\\
A_0(k_i,z) &= a_0(k_i) I(\vec{k},z) + b_0(k_i) K(\vec{k},z), \qquad 
\vec{k} \equiv (\delta_{ij}k^i k^j)^{1/2},
\end{align}
where the two modes
\begin{align}
I(\vec{k},z) &= \frac{\Gamma(d/2)}{d-2}\left(\frac{2 z}{\vec{k}}\right)^{d/2-1} I_{d/2-1}(\vec{k}z), \qquad 
K(\vec{k},z) = \frac{2(\vec{k}z/2)^{d/2-1}}{\Gamma(\frac{d}{2}-1)} K_{d/2-1}(\vec{k}z), 
\end{align}
are normalized such that for small $z$ we have $I(\vec{k}z)\sim z^{d-2}/(d-2)$ and $K(\vec{k}z)\sim 1$. 
To obtain the Green's function we solve
\begin{equation}
z^{d+1} \partial_z (z^{3-d} \partial_z \Phi(\vec{k}; z,z')) - z^{4} \vec{k}^2 \Phi(\vec{k}; z,z') = -z^{d+1}\delta(z-z'),
\end{equation}
with the boundary condition that $\Phi=0$ at $z=0$ and $z=\infty$.
The unique solution is
\begin{align}
\Phi(\vec{k}; z,z') =
K(\vec{k},z') I(\vec{k},z)\Theta(z'-z)  +
K(\vec{k},z) I(\vec{k},z')\Theta(z-z'),
%\implies \Phi(u,u') = \int \frac{d^{d-1}k}{(2\pi)^{d-1}}e^{i k^i(x_i-x'_i)} \tilde \Phi(k;z,z').
\end{align}
Fourier transforming back and integrating with the current operator to get the general solution as before, we find 
%\begin{equation}
%A_0(t,u) = \int \frac{d^du}{z^{d+1}} \Phi(u, u') J^0(t,u')
%\end{equation}
%and therefore
\begin{equation}
[\hat A_0(t,u), \hat\phi(t,u')] = q \Phi(u,u') \hat\phi(t,u').
\end{equation}
Taking the electric field to the boundary and applying the dictionary \eqref{dictionary}, we get
\begin{equation} \label{currentscalarcommutator}
[\hat j^0(x),\hat \phi(x',z')]
%\lim_{z\to 0} z^{3-d}[F_{z0}, \phi] 
%= \lim_{z\to 0} z^{3-d} \partial_z [\hat A_0(t,u), \hat\phi(t,u')] 
= q \hat\phi(t,u') \int \frac{d^{d-1}k}{(2\pi)^{d-1}}e^{i k_jx^j} K(\vec{k},z').
\end{equation}
We have a single trace local operator in the CFT that does not commute with the bulk scalar field.
Again this can be understood as a consequence of charge conservation and \eqref{currentscalarcommutator} can be integrated over a spatial slice through the boundary to yield the commutator of the total charge with the field.
%$$ [\hat Q, \hat\phi] = q \hat\phi.$$
% \begin{equation}
% \int d^{d-1}x [\hat j^0(x),\hat \phi(x,z)] =
% \end{equation}

With the bulk dimension greater than two there are photons and from the mode expansion we can get an expression for the field operator in terms of the boundary current, analogous to \eqref{smearingscalar}. 
For timelike boundary momenta $k^2=\eta^{\mu\nu}k_\mu k_\nu <0$ the mode solutions to \eqref{spatialMaxwelldecoupled}  are
\begin{align}
A_i &= a_i(k_\mu)F(k,z)+ b_i(k_\mu)Y(k,z), 
\qquad
A_z = a_z(k_\mu) F_z(k,z)+ b_z(k_\mu) Y_z(k,z), 
\end{align}
where we have defined
\begin{equation}
F(k,z) =  \frac{\Gamma(\frac{d}{2})}{d-2}\left(\frac{2z}{|k|}\right)^{\frac{d}{2}-1} J_{\frac{d}{2}-1}(|k|z), \qquad
F_z(k,z) =  \Gamma(\frac{d}{2}-1)\left(\frac{2z}{|k|}\right)^{\frac{d}{2}-1} J_{\frac{d}{2}-2}(|k|z).
\end{equation}
and similarly $Y$ and $Y_z$ with the substitution $J_\nu \to Y_\nu$. The normalization is such that near the bounary we have 
$$\partial_z F \sim F_z\sim  z^{d-3}.$$ 
For the solutions $Y_a$ normalization does not matter as we now impose the standard boundary condition $b_a=0$.
From the dictionary \eqref{dictionary} for timelike boundary momentum we then find
\begin{equation}
j_i(k_\mu) = a_i(k_\mu) - k_i a_z(k_\mu).
\end{equation}
Substituting the solution into the gauge condition we get $k_j a_j(k_\mu) = 0$: the spatial components along the boundary are transverse with respect to the spatial momentum along the boundary.
Therefore we identify $k_i a_z$ with the longitudinal part of the boundary current and $a_i$ with the transverse part
\begin{equation}
a_i (k_\mu) =  \left(\delta_{ij} - \frac{k_ik_j}{\vec{k}^2}\right) j_j(k_\mu), \qquad 
a_z = \frac{k_i j_i(k_\mu)}{\vec{k}^2}.
\end{equation}
% This then leads to the operator expression
% \begin{align}
% \hat A_i(z,x) &= \int_{k^0 > |k|^2} \frac{d^dk}{(2\pi)^d} \left(\delta_{ij} - \frac{k_ik_j}{\vec{k}^2}\right) \hat j_j(k_\mu) J(k,z) e^{i k_\mu x^\mu}  + h.c., 
% \\
% \hat A_z(z,x) &= \int_{k^0 > |k|^2} \frac{d^dk}{(2\pi)^d} \frac{k_i\hat j_i(k_\mu)}{\vec{k}^2} J_z(k,z)e^{i k_\mu x^\mu} + h.c.
% \end{align}
When we plug this into the mode expansion, we obtain after inverse Fourier transforming $j_i(k_\mu)$ and swapping the integrals
\begin{align} \label{Asmearing}
\hat A_a(z,x) &= \int d^dx' K^{(0)}_{aj}(z; x-x') \hat j_j(x'), 
\end{align}
where the leading order smearing functions are
\begin{align}
K^{(0)}_{ij}(z,x-x') &= \mathrm{Re} \int_{k^0 > \vec{k}} \frac{d^dk}{(2\pi)^d} \left(\delta_{ij} - \frac{k_ik_j}{\vec{k}^2}\right) F(k,z) e^{i k_\mu (x^\mu-x'^\mu)}  
\\
K^{(0)}_{zj}(z,x-x') &= \mathrm{Re}\int_{k^0 > \vec{k}} \frac{d^dk}{(2\pi)^d} \frac{k_j}{\vec{k}^2} F_z(k,z) e^{i k_\mu (x^\mu-x'^\mu)}  
\end{align}
The restriction of the integration range to timelike momenta is because from Lorentz invariance $\hat j^\mu(k_\mu) = 0$ for spacelike momenta.

We can also easily obtain the smearing function in radial gauge $A_z=0$, where the Maxwell equations take the form
\begin{align}
z^{d+1} \partial_z (z^{3-d} \partial_z A^\mu) + z^{4} \partial^\rho(\partial_\rho A^\mu - \partial^\mu A_\rho) = J^\mu 
,\qquad
-z^4 \partial_z \partial_\mu A^\mu = J^z
\end{align}
If we impose the supplementary gauge condition that $\partial_\mu A^\mu=0$ on the boundary\footnote{Alternatively, and from some perspective more naturally, we can fix the non-normalizable mode without imposing transversality on the boundary and transversality for the normalizable mode will be implied by the equation of motion.}, and solve the sourceless equations of motion, we find the general solution
\begin{align}
A^\mu(z,k) = a_\mu (k_\mu) F(k,z) + b_\mu(k_\mu) Y(k,z), \qquad k^\mu a_\mu = k^\mu b_\mu = 0.
\end{align}
Setting $b_\mu=0$ we find from the dictionary that $a_\mu = j_\mu$ so that
\begin{equation}
\hat A_\mu(z,x) = \int d^dx' K_{rad}^{(0)}(z,x-x') \hat j_\mu(x'),
\end{equation}
with the simple smearing function
\begin{equation}
K^{(0)}_{rad}(z,x-x') = \mathrm{Re} \int_{k^0 > \vec{k}} \frac{d^dk}{(2\pi)^d}  F(k,z)e^{ik_\mu (x^\mu-x'^\mu)}.
\end{equation}

We could have obtained the same smearing functions by solving for the Green's functions for the spatial equations and applying Green's theorem. For finding the results in this section that method is less direct but in general it makes it possible to find nicer smearing functions.

\section{Linearized Einstein gravity} \label{gravitysection}
Although there is significantly more gauge redundancy in the case of gravity, perturbatively the structure is similar to electromagnetism. We will consider gravity coupled to a free massless scalar
\begin{equation}
S = \frac{1}{2\kappa}\int d^{d+1}y (R - 2\Lambda) - \frac{1}{2}\int d^{d+1}y g^{MN} \partial_M \phi \partial_N \phi
\end{equation}
with $\Lambda = -d(d-1)/2$ 
and work to linearized order in metric perturbations around the AdS background 
$$g_{MN} = \bar g_{MN} + h_{MN}.$$
Since gravity is trivial in $d=1$ we will immediately work in arbitrary dimension $d>1$.
The expansion of the Einstein-Hilbert action with cosmological constant to second order in $h_{MN}$ is given in \cite{Liu:1998bu},  and from it we obtain the linearized Einstein equations on AdS, which can be written as
\begin{align} \label{linearEinstein}
2\kappa T_{MN}
=
\bar \nabla_{K} \bar \nabla_{(M} \tilde h^K_{\hspace{5pt} N)}
 + \nabla_{(M}\nabla^K \tilde h_{N)K} 
- \bar \nabla^2 \tilde h_{MN} 
-\bar \nabla_K \bar \nabla^L \tilde h^K_{\hspace{5pt} L} &\bar g_{MN} 
\\\nonumber
&+ \tilde h g_{MN}+ (d-1) \tilde h_{MN}.
\end{align}
Here $\tilde h_{MN} = h_{MN}- \frac{1}{2} h \bar g_{MN}$, known as the trace reversed metric in four dimensions. The stress tensor for the scalar field is given by
\begin{equation} \label{scalarstresstensor}
T_{MN} = \nabla_M \phi \nabla_N \phi - \frac{1}{2} (\nabla\phi)^2 g_{MN}.
\end{equation}
and using the canonical commutators we find 
\begin{equation}
[T_{0M}(t,u), \phi(t,u')] = -iz^{d-1} \delta^d(u-u') \partial'_M \phi(u').
\end{equation}

\subsection{Gauss' law and holography}
To solve the gravitational Gauss' law and see that it leads to a non-local commutator it is again easiest to work in transverse gauge. While this simplifies Gauss' law, which in this context we will refer to as the Hamiltonian constraint, the other equations are hard to solve and the AdS/CFT dictionary is messy. Therefore, we will not work out the full canonical quantization and smearing functions in this gauge but only discuss the gravitational Gauss' law.
To generalize the flat space transverse gauge we first decompose the metric as follows
\begin{equation}
h_{00} = 2\Phi \bar g_{00}, \qquad h_{0a} = w_a \bar n_0, \qquad h_{ab} = 2s_{ab} - 2\Psi \bar g_{ab}, \qquad s_{ab}\bar g^{ab}=0,
\end{equation}
with $\bar n^M$ the unit timelike normal in the background.
The Hamiltonian constraint (the 00 component of \eqref{linearEinstein}) now takes the form
\begin{align}
\kappa T_{00}\bar g^{00}
&=
-(d-1) \bar \nabla^a \bar \nabla_a \Psi - \bar\nabla_a \bar\nabla_b s^{ab} + (d-1)d \Psi,
\end{align}
and after imposing the transverse gauge condition
\begin{equation}
\bar\nabla_b s^{ab} = 0, 
\end{equation}
it can be written as a Poisson type equation
\begin{align} \label{GraviGauss}
-z^{d-1} \partial_a (z^{-d} \partial_a (z \Psi)) = \frac{\kappa}{d-1} T_{00},
\end{align}
which is solved by 
\begin{align}
\Psi(t,u) = \frac{\kappa}{d-1}\int d^du' z'^{1-d} \Phi_G(u,u') T_{00}(t,u'),
\end{align}
where
\begin{align}
\Phi_G(\vec{k}; z,z') =
z' K_G(\vec{k},z') I_G(\vec{k},z)\Theta(z'-z)  +
z' K_G(\vec{k},z) I_G(\vec{k},z')\Theta(z-z')
%\implies \Phi(u,u') = \int \frac{d^{d-1}k}{(2\pi)^{d-1}}e^{i k^i(x_i-x'_i)} \tilde \Phi(k;z,z').
\end{align}
with the two modes 
$$I_G(\vec{k},z) \propto z^{(d-1)/2} I_{(d+1)/2}(\vec{k}z),
\qquad K_G(\vec{k},z) \propto z^{(d-1)/2} K_{(d+1)/2}(\vec{k}z),$$
normalized such that near the boundary we have $\partial_z (z I_G(\vec{k},z)) \sim z^d$ and $K_G(\vec{k},z)\sim 1/z$.
Just as in the electromagnetic case we find a non-local commutator
\begin{equation} \label{gravcommutator}
[\Psi(t,u), \phi(t,u')] = -i \Phi_G(u,u') \partial_0 \phi(u').
\end{equation}
To first order in the perturbation we have a timelike Killing field $\xi^N = \delta^N_0$ so that to this order we have a conserved energy
\begin{equation}
H = \int d^d u \sqrt{\gamma} \, \bar n^M \xi^N T_{MN} = \int \frac{d^d u}{z^{d-1}} T_{00}.
\end{equation}
Applying this to \eqref{GraviGauss} we obtain
\begin{align} \label{Hbdryterm}
H = \frac{d-1}{\kappa} \lim_{z\to 0} \int_{\partial M} d^d x z^{-d} \partial_z (z\Psi).
\end{align}
while applying it to \eqref{gravcommutator} we find, as we should $[H,\phi]=-i\partial_0\phi$.

All this is exactly analogous to the gauge field. However, for gravity it has profound consequences because the conserved charge is the time evolution operator. 
As explained in \cite{arXiv:0808.2842,arXiv:0808.2845} and made explicit by the smearing function construction of bulk operators, the boundary values of the fluctuating modes of the bulk fields are a complete set of observables. If we know the boundary values at all times we can reconstruct all the bulk operators.
From \eqref{Hbdryterm} we see that the bulk Hamiltonian is a pure boundary term: it only depends on the boundary value of $\Psi$. What this means is that in fact the boundary values at a single time are a complete set, because from the commutators with the boundary value of $\Psi$ with the other boundary values at fixed time we can obtain the boundary values at all times. This gives us a perturbative explanation of holography in the sense that while the number of degrees of freedom is not restricted, we see that at fixed time all bulk information is contained in the boundary.

\subsection{Smearing function in radial gauge}
For the source free case \eqref{linearEinstein} can be written as
\begin{align} \label{homogeneousgravity}
2 \bar \nabla_{K} \bar \nabla_{(M} h^K_{\hspace{4pt} N)} 
- \bar \nabla_{M} \bar \nabla_{N} h  
- \bar \nabla^2 h_{MN} + 2d h_{MN}  =0.
\end{align}
As for the gauge field, finding a smearing function is much easier in radial (Pfefferman-Graham ) gauge $h_{zM}=0$. From \cite{Compere:2008us} we find that, analogous to the electromagnetic case, we can impose supplementary gauge conditions $\partial_\mu h^{\mu \nu} = 0$ and $\eta^{\mu\nu}h_{\mu\nu}=0$ on the boundary after which the equations imply that these conditions hold everywhere.
Now \eqref{homogeneousgravity} reduces to 
\begin{align}
z^2 \partial_z^2 h_{\mu\nu} + (5-d) z \partial_z h_{\mu\nu} + z^2 \partial^2 h_{\mu\nu} - 2(d-2) h_{\mu\nu} = 0,
\end{align}
which is solved by
\begin{align}
h_{\mu\nu} = h_{(d)\mu\nu}(k_\mu) F_G(k,z) + h_{(0)\mu\nu} Y_G(k,z),
\end{align}
where $F_G(z,k) \propto z^{d/2-2} J_{d/2}(kz)$ is normalized on the boundary and $Y_G$ is defined similarly with $J_\nu$ replaced by $Y_\nu$. In standard quantization we impose the boundary condition $h_{(0)\mu\nu}=0$ and the dictionary takes the form \cite{de Haro:2000xn}
\begin{align}
\frac{d}{2\kappa} \hat h_{(d)\mu\nu} = \hat T^{CFT}_{\mu\nu}.
\end{align}
We therefore find
\begin{align}
\hat h_{\mu\nu}(z,x) = \int d^dx' K^{(0}_G(z, x-x') \hat T_{\mu\nu}^{CFT}(x'),
\end{align}
with the smearing function
\begin{align}
K^{(0}_G(z, x-x') = \frac{2\kappa}{d}\mathrm{Re} \int_{k_0>\vec{k}} \frac{d^dk}{(2\pi)^d} F_G(k,z) e^{ik_\mu (x^\mu-x'^\mu)}.
\end{align}

\section{Discussion}

It would be nice to find covariant smearing functions of spacelike support using a Green's function method, analogous to what was done in the scalar case in \cite{Hamilton:2006az, Hamilton:2005ju}. However there is no way to completely fix the gauge in a covariant way. 
Perhaps this is not a problem if instead one tried to find smearing functions for gauge invariant operators in the bulk such as the field strength, instead of the gauge field.
It would also be very interesting to understand the non-perturbative breakdown of the construction.

%  At finite $N$ trace relations between boundary operators will limit locality in the bulk. To consider this while still having a classical bulk we imagine building up all the component fields in the bulk string field theory as smeared boundary operators. Then for two heavy operators
% \begin{align}
% \Op_1 = \mathrm{Tr}(X^N), \qquad \Op_2 = \mathrm{Tr}(X^3)\mathrm{Tr}(X^{N-3})
% \end{align}
%Non-perturbative non-locality? Difference between trace relations and $e^{-N^2(t'-t')}$ type corrections. How do NP effects show up in Heisenberg equation? Like perturbative interactions you can't neglect them in the bulk and smear a NP bdry correlator.

\section*{Acknowledgements}
I thank Joe Polchinski for being a great advisor and Jamie Sully, Ahmed Almuhairi, Tomas Andrade, Jorge Santos, Ian Morrison and especially Don Marolf for discussions.
This work was supported in part by NSF grant PHY07-57035 and FQXi grant RFP3-1017.


\begin{thebibliography}{99}
\bibitem{BDHM}
  T.~Banks, M.~R.~Douglas, G.~T.~Horowitz and E.~J.~Martinec,
  ``AdS dynamics from conformal field theory,''
  arXiv:hep-th/9808016.
  %%CITATION = HEP-TH/9808016;%%

\bibitem{Bena99}
  I.~Bena,
  ``On the construction of local fields in the bulk of AdS(5) and other spaces,''
  Phys.\ Rev.\  {\bf D62}, 066007 (2000).
  [hep-th/9905186].

\bibitem{KLL2011}
  D.~Kabat, G.~Lifschytz, D.~A.~Lowe,
  ``Constructing local bulk observables in interacting AdS/CFT,''
  Phys.\ Rev.\  {\bf D83}, 106009 (2011).
  [arXiv:1102.2910 [hep-th]].

\bibitem{ourotherpaper}
  I.~Heemskerk, D.~Marolf, J.~Polchinski,
 ``Bulk and Transhorizon Measurements in AdS/CFT',''
 [arXiv:1201.xxxx [hep-th]]

\bibitem{Gary:2009ae}
  M.~Gary, S.~B.~Giddings and J.~Penedones,
  ``Local bulk S-matrix elements and CFT singularities,''
  Phys.\ Rev.\  D {\bf 80}, 085005 (2009)
  [arXiv:0903.4437 [hep-th]].
  %%CITATION = PHRVA,D80,085005;%%

\bibitem{HoloFromCFT}
  I.~Heemskerk, J.~Penedones, J.~Polchinski, J.~Sully,
  ``Holography from Conformal Field Theory,''
  JHEP {\bf 0910}, 079 (2009).
  [arXiv:0907.0151 [hep-th]].

%\cite{Heemskerk:2010hk}
\bibitem{WHRG}
  I.~Heemskerk and J.~Polchinski,
  ``Holographic and Wilsonian Renormalization Groups,''
  JHEP {\bf 1106}, 031 (2011)
  [arXiv:1010.1264 [hep-th]].

%\cite{Nguyen:1989iz}
\bibitem{Nguyen:1989iz}
  S.~H.~Nguyen and V.~N.~Pervushin,
  ``Gauge-Invariant Quantization Of Abelian And Non-Abelian Theories,''
  Fortsch.\ Phys.\  {\bf 37}, 611 (1989).
  %%CITATION = FPYKA,37,611;%%

%\cite{873163}
\bibitem{BjorkenDrell} 
  J.~D.~Bjorken and S.~D.~Drell,
  ``Relativistic quantum fields,''
  %%CITATION = INSPIRE-873163;%%

%\cite{Polchinski:2010hw}
\bibitem{Polchinski:2010hw}
  J.~Polchinski,
  ``Introduction to Gauge/Gravity Duality,''
  arXiv:1010.6134 [hep-th].
  %%CITATION = ARXIV:1010.6134;%%

%\cite{Liu:1998bu}
\bibitem{Liu:1998bu} 
  H.~Liu and A.~A.~Tseytlin,
  ``D = 4 superYang-Mills, D = 5 gauged supergravity, and D = 4 conformal supergravity,''
  Nucl.\ Phys.\ B {\bf 533}, 88 (1998)
  [hep-th/9804083].
  %%CITATION = HEP-TH/9804083;%%

%\cite{Compere:2008us}
\bibitem{Compere:2008us}
  G.~Compere and D.~Marolf,
  ``Setting the boundary free in AdS/CFT,''
  Class.\ Quant.\ Grav.\  {\bf 25}, 195014 (2008)
  [arXiv:0805.1902 [hep-th]].
  %%CITATION = CQGRD,25,195014;%%

%\cite{de Haro:2000xn}
\bibitem{de Haro:2000xn} 
  S.~de Haro, S.~N.~Solodukhin and K.~Skenderis,
  ``Holographic reconstruction of space-time and renormalization in the AdS / CFT correspondence,''
  Commun.\ Math.\ Phys.\  {\bf 217}, 595 (2001)
  [hep-th/0002230].
  %%CITATION = HEP-TH/0002230;%%

%\cite{arXiv:0808.2842}
\bibitem{arXiv:0808.2842} 
  D.~Marolf,
  ``Unitarity and Holography in Gravitational Physics,''
  Phys.\ Rev.\ D\ {\bf 79}, 044010  (2009)
  [arXiv:0808.2842 [gr-qc]].
  %%CITATION = PHRVA,D79,044010;%%

%\cite{arXiv:0808.2845}
\bibitem{arXiv:0808.2845} 
  D.~Marolf,
  ``Holographic Thought Experiments,''
  Phys.\ Rev.\ D\ {\bf 79}, 024029  (2009)
  [arXiv:0808.2845 [gr-qc]].
  %%CITATION = PHRVA,D79,024029;%%

%\cite{Hamilton:2005ju}
\bibitem{Hamilton:2005ju}
  A.~Hamilton, D.~N.~Kabat, G.~Lifschytz and D.~A.~Lowe,
  ``Local bulk operators in AdS/CFT: A boundary view of horizons and
 locality,''
  Phys.\ Rev.\  D {\bf 73}, 086003 (2006)
  [arXiv:hep-th/0506118].
  %%CITATION = PHRVA,D73,086003;%%

%\cite{Hamilton:2006az}
\bibitem{Hamilton:2006az}
  A.~Hamilton, D.~N.~Kabat, G.~Lifschytz and D.~A.~Lowe,
  ``Holographic representation of local bulk operators,''
  Phys.\ Rev.\  D {\bf 74}, 066009 (2006)
  [arXiv:hep-th/0606141].
  %%CITATION = PHRVA,D74,066009;%%

\end{thebibliography}
\end{document}